\documentclass[ALICE,manyauthors]{cernphprep}
\usepackage[english]{babel}
\usepackage{graphicx}
\usepackage{verbatim}
\usepackage{amsfonts}
\usepackage{makeidx}
\usepackage{color}
\usepackage{amsmath}
\usepackage{lineno}
\usepackage{epsfig}
\usepackage{epstopdf}
\usepackage{hyperref}
\usepackage{cite}

\newcommand{\pp}{pp}
\newcommand{\mpp}{\mathrm{pp}}
\newcommand{\sqrts}{\sqrt{s}}

\newcommand{\mev}{\mathrm{MeV}}
\newcommand{\gev}{\mathrm{GeV}}
\newcommand{\gevc}{\mathrm{GeV}/c}
\newcommand{\tev}{\mathrm{TeV}}

\newcommand{\mum}{\mathrm{\mu m}}

\newcommand{\mb}{\mathrm{mb}}
\newcommand{\mub}{\mathrm{\mu b}}

\newcommand{\PbPb}{\mbox{Pb--Pb}}
\newcommand{\pt}{p_{\rm t}}

\newcommand{\DtoKpi}{{\rm D}^0 \to {\rm K}^-\pi^+}
\newcommand{\DtoKpipi}{{\rm D}^+\to {\rm K}^-\pi^+\pi^+}
\newcommand{\DstartoDpi}{{\rm D}^{*+} \to {\rm D}^0 \pi^+}
\newcommand{\DstartoDpiPrecise}{{\rm D^{*+}(2010)\to D^0\pi^+}}
\newcommand{\Dzero}{{\rm D^0}}
\newcommand{\Dzerobar}{\overline{{\rm D^0}}}
\newcommand{\Dstar}{{\rm D^{*+}}}
\newcommand{\Dstarm}{{\rm D^{*-}}}
\newcommand{\Dplus}{{\rm D^+}}
\newcommand{\Dminus}{{\rm D^-}}
\newcommand{\ccbar}{${\rm c}\bar{\rm c}$~}
\newcommand{\nbinv}{{\rm nb^{-1}}}

\newcommand{\dEdx}{{\rm d}E/{\rm d}x}
\newcommand{\Pv}{{\rm P}_{\rm v}}

\newcommand{\cdbar}{{\rm c}\bar{\rm d}}
\newcommand{\mur}{\mu_{\rm R}}
\newcommand{\muf}{\mu_{\rm F}}
\newcommand{\mc}{m_{\rm c}}
\newcommand{\sigmatot}{\sigma_{\rm tot}}

\begin{document}

\begin{titlepage}

\PHnumber{2012-133}                 
\PHdate{17 May 2012}              

\title{Measurement of charm production at central rapidity in proton--proton collisions at $\sqrts=2.76$~TeV}

\Collaboration{ALICE Collaboration%
         \thanks{See Appendix~\ref{app:collab} for the list of collaboration 
                      members}}
\ShortAuthor{ALICE Collaboration}
\ShortTitle{Charm production at central rapidity in pp at $\sqrts=2.76$~TeV}

\begin{abstract} 
The $\pt$-differential production cross sections of the prompt (B feed-down subtracted) charmed mesons $\Dzero$, $\Dplus$, and $\Dstar$ in the rapidity range $|y|<0.5$, and for transverse momentum $1< \pt <12$~$\gevc$, were measured in proton--proton collisions at $\sqrt{s}=2.76~\tev$ with the ALICE detector at the Large Hadron Collider. The analysis exploited the hadronic decays $\DtoKpi$, $\DtoKpipi$, $\DstartoDpi$, and their charge conjugates, and was performed on a $\mathcal{L}_{{\rm int}}=1.1$~$\nbinv$ event sample collected in 2011 with a minimum-bias trigger. 
The total charm production cross section at $\sqrt{s}=2.76~\tev$ and at $7~\tev$ was evaluated by extrapolating to the full phase space the $\pt$-differential production cross sections at $\sqrt{s}=2.76~\tev$ and our previous measurements at $\sqrts=7~\tev$. 
The results were compared to existing measurements and to perturbative-QCD calculations. 
The fraction of $\cdbar$ D mesons produced in a vector state was also determined. 
\end{abstract}


\end{titlepage}
\setcounter{page}{2}

\section{Introduction}
\label{sec:intro}
The measurement of charm and beauty production cross sections in proton--proton (\pp) collisions at the Large Hadron Collider (LHC) constitutes an important test of perturbative Quantum Chromo--Dynamic (pQCD) calculations at the highest available collider energies.  
These calculations use the factorization approach to describe heavy-flavour hadron production as a convolution of three terms: the parton distribution function, the hard parton scattering cross section and the fragmentation function. The parton distribution function describes the initial distribution of quarks and gluons from the colliding protons. The hard parton scattering cross section is calculated as a perturbative series in the coupling constant of strong interaction. The fragmentation function parametrizes the relative production yield and momentum distribution for a heavy quark that hadronizes to particular hadron species.
The production cross section of beauty hadrons at Tevatron ($\sqrts=1.96~\tev$)~\cite{cdfB,fonllBcdf,gmvfnsBcdf} and at the LHC ($\sqrts=7~\tev$)~\cite{lhcbBeauty,cmsJpsi} is well described by perturbative calculations at next-to-leading order (e.g. GM-VFNS~\cite{gmvfns}) or at fixed order with next-to-leading-log resummation (FONLL~\cite{fonll}). 
The production cross section of charmed hadrons at Tevatron~\cite{charmcdf,fonllDcdf,gmvfnsDcdf} and at the LHC~\cite{aliceDmeson7TeV,ATLASDmeson,LHCbDmeson}, as well as the RHIC heavy-flavour decay lepton measurements at  $\sqrts=200~\gev$~\cite{phenixelepp,starelepp}, are also well reproduced within the uncertainties of the pQCD calculations. 
However, the overall comparison suggests that the calculation, as obtained with its central parameters, underestimates charm production. 
The measurement of charm production as a function of the centre-of-mass energy therefore provides an interesting probe of pQCD. 
The relative abundances of open charmed hadrons also test the statistical hadronization scenario~\cite{shm-charm} of charm quarks into hadrons, which then should be independent of the collision system and energy. 
Finally, heavy quarks provide a unique probe for studies of the properties of the QCD matter created in \PbPb~collisions at unprecedented high energies at the LHC (see e.g.~\cite{stathadronization,hqcolorimetry}). Heavy-quark production rates in \pp~collisions provide the necessary baseline for such studies and motivates the measurement reported in this paper.

We present the measurement of the production cross section of the prompt (B feed-down subtracted) charmed mesons $\Dzero$, $\Dplus$, and $\Dstar$, in \pp~collisions at $\sqrts=2.76~\tev$ in $|y|< 0.5$, reconstructed in the range $2< \pt <12$~$\gevc$ ($1< \pt <12$~$\gevc$ for the $\Dzero$) with the ALICE experiment~\cite{aliceJINST}. The apparatus is described in section~\ref{sec:detector}, along with the data sample used for the measurement. The D meson analysis (reconstruction, signal extraction, corrections, systematic uncertainties) is presented in section~\ref{sec:DataAnalysis}. The $\pt$-differential cross sections are reported in section~\ref{sec:crosssection}, and are compared to theoretical QCD calculations and to the ALICE measurements at $\sqrts=7~\tev$~\cite{aliceDmeson7TeV} scaled to $\sqrts=2.76~\tev$ by a pQCD-driven scaling~\cite{EnergyExtrap}. 
In section~\ref{sec:charmAndPv}, the visible cross sections at $\sqrts=2.76~\tev$ and $\sqrts=7~\tev$ are extrapolated to the full phase space to calculate the fraction of $\cdbar$~D mesons produced in a vector state and the total \ccbar production cross section at these two energies. 
The results are compared with existing measurements and predictions.

\section{Experimental apparatus and data sample}
\label{sec:detector}
The D mesons are reconstructed in the central rapidity region using the central barrel detectors of the ALICE experiment.
In the following, the detectors utilized for the D meson analysis are discussed. A detailed description of the ALICE apparatus is given in Ref.~\cite{aliceJINST}. 
The central barrel detectors are contained in a large solenoidal magnet, which provides a magnetic field of 0.5~T along the beam direction. 
The closest detector to the beam axis ($z$) is the Inner Tracking System (ITS). It is made of six cylindrical layers of silicon detectors with radii between 3.9 and 43.0~cm. The two innermost layers, with radii 3.9~cm (0.9~cm from the beam vacuum tube) and 7.6~cm, are equipped with Silicon Pixel Detectors (SPD). The two intermediate layers (at radii of 15.0 and 23.9~cm) are made of Silicon Drift Detectors (SDD). Finally the two outermost layers are equipped with Silicon Strip Detectors (SSD) and are located at radii of 38.0 and 43.0~cm. 
The total material budget of the ITS is on average 7.7\% of a radiation length for charged particles crossing the ITS perpendicularly to the detector surfaces $(\eta = 0)$~\cite{aliceJINST,ITSalign}. 
The experiment's low magnetic field allows to track low $\pt$ hadrons (about 80~$\mev/c$ for pions). 
In the present analysis, the main role of the ITS was to resolve the topology of the hadronic decays of the D mesons by identifying the secondary vertex of the decay. 
A particularity of the data sample collected in \pp~collisions at $\sqrts=2.76$~TeV was that, in order to collect a higher statistics data sample, minimum-bias events were triggered independently of the SDD read-out state. This resulted in a fraction of events missing the SDD information. 
To have a homogeneously reconstructed sample of tracks, the SDD points were always excluded from the track reconstruction used for this analysis. 

The ITS is surrounded by a 510~cm long cylindrical Time Projection Chamber (TPC) that covers $|\eta| < 0.9$~\cite{TPC}. It provides track reconstruction with up to 159 points along the trajectory of a charged particle, as well as particle identification via specific energy deposit $\dEdx$. The Time-Of-Flight (TOF) detector, based on Multi-gap Resistive Plate Chambers (MRPCs), is positioned in the region between 377 to 399~cm from the beam axis and covers $|\eta| < 0.9$ and full azimuth. In this analysis, the TOF complemented the hadron identification capability of the TPC, ensuring an efficient track by track kaon/pion separation up to a momentum of about 1.5~$\gevc$. With the present level of calibration, the intrinsic timing resolution was better than 100~ps. 
The overall TOF resolution including the uncertainty on the start time of the event, which is the time at which the collision took place, and the tracking and momentum resolution contributions, was, on average, around 150~ps~\cite{TOFcal}. 
The event start time information was provided by the T0 detector. It is formed by two arrays of 12 Cherenkov counters each, one located at $-3.28< \eta < -2.97$, at 72.7~cm from the interaction point (IP), and the other at $4.61< \eta < 4.92$, at 375~cm from the IP~\cite{aliceJINST}. 
The event start time information was also estimated using the particle arrival times at the TOF detector. This was particularly useful for the events in which the T0 signal was not present. 
For the events where the number of tracks was not sufficient to apply this method, and, at the same time, there was no information from the T0 detector, the bunch-crossing time from the LHC was used as the event start time. 
%

Minimum-bias collisions were triggered by requiring at least one hit in either of the VZERO scintillator hodoscopes (one located at $z=328$~cm covering $2.8<\eta<5.1$, and the other at $z=-86$~cm covering $-3.7<\eta<-1.7$) or in the SPD ($|\eta|<2$), in coincidence with the arrival of proton bunches from both sides of the interaction region. 
This trigger configuration was estimated to be sensitive to about $87\%$ of the \pp~inelastic cross section~\cite{oyama}. PYTHIA 6.4.21~\cite{pythia} Monte Carlo simulations (with Perugia-0 tune~\cite{perugia0}), using GEANT3~\cite{geant3} and including the description of the detector geometry, material and response, confirmed that this minimum-bias trigger is 100\% efficient for D mesons with $\pt > 1$~GeV/$c$ and $|y| < 0.5$. Contamination from beam-induced background interactions was rejected offline using the timing information from the VZERO detector and the correlation between the number of hits and track segments (tracklets) in the SPD detector. The probability of collision pile-up per triggered event was kept below 2.5\% by limiting the instantaneous luminosity in the ALICE experiment to $4.9 \times 10^{29}$~cm$^2$s$^{-1}$. 
The pile-up events were tagged as those where two interaction vertices, separated by more than 8~mm, and having at least 3 associated tracklets (hit pairs in the two layers of the SPD), were found. The remaining pile-up events, less than four per mille, were negligible in the present analysis. 
The Gaussian r.m.s. of the interaction region was measured to be $\sigma_{x} \approx \sigma_{y} \approx 100~\mu$m and $\sigma_{z} \approx 5.5$~cm from the distribution of the interaction vertices reconstructed with the charged particles tracked in the central detectors. 
The resolution of the primary vertex position depends on the charged particle multiplicity (${\rm d}N_{\rm ch}/{\rm d}\eta$). It ranges within $100~\mu$m, for ${\rm d}N_{\rm ch}/{\rm d}\eta<5$, and $20~\mu$m, for ${\rm d}N_{\rm ch}/{\rm d}\eta \sim 30$. 
Only events with no pile-up and a reconstructed vertex within $|z|< 10$~cm from the centre of the detector were kept, resulting in $58$~M events analyzed, corresponding to an integrated luminosity $\mathcal{L}_{int}=1.1$~$\nbinv$. The integrated luminosity was evaluated as $\mathcal{L}_{int} = N_{\mpp,{\rm MB}} / \sigma_{\mpp,{\rm MB}}$, where $N_{\mpp,{\rm MB}}$ and $\sigma_{\mpp,{\rm MB}}$ are the number and cross section of \pp~collisions passing the minimum-bias trigger condition. The value of $\sigma_{\mpp,{\rm MB}}=54.8$~mb was determined from the measurement of the \pp~collisions that gave signals in both sides of the VZERO scintillator detector using a van der Meer scan ($\sigma_{\mpp,{\rm VZERO-AND}}$)~\cite{oyama}. The normalization factor, $\sigma_{\mpp,{\rm VZERO-AND}}/\sigma_{\mpp,{\rm MB}} \approx 0.87$, was found to be stable within 1\% in the data sample. The systematic uncertainty of 1.9\% was assigned upon considering the uncertainties on the beam intensities and on the analysis procedure.

\section{D meson analysis}
\label{sec:DataAnalysis}

\subsection{D meson reconstruction and selection}
\label{sec:signal}
The study of charm production was performed by reconstructing $\Dzero$, $\Dplus$, and $\Dstar$ charmed hadrons via their hadronic decays $\DtoKpi$ (BR of $3.87\pm 0.05\%$~\cite{pdg}), $\DtoKpipi$ (BR of $9.13\pm 0.19\%$~\cite{pdg}, including the resonant channels via a $\rm K^{*0}$), and $\DstartoDpiPrecise$  (BR of $67.7\pm 0.5\%$~\cite{pdg}) with $\DtoKpi$, and their charge conjugates. 
The decays of $\Dzero$ and $\Dplus$ are weak processes with mean proper decay lengths $c\tau\approx 123$ and $312~\mum$. Their secondary decay vertices are then typically displaced by a few hundred $\mum$ from the primary interaction vertex.  
The analysis strategy for the $\Dzero$ and $\Dplus$ was based on the reconstruction and selection of secondary vertex topologies with significant separation from the primary vertex. The topological reconstruction of the decay allowed for an efficient rejection of the combinatorial background from uncorrelated tracks. The identification of the charged kaon using the TPC and the TOF detectors provided additional background rejection in the low-momentum region. 
Finally, the signal was extracted by an invariant mass analysis of the candidate pairs and triplets.
In the $\Dstar$ case, since the decay proceeds via the strong interaction, it is not possible to resolve the secondary $\Dstar$ vertex. The analysis exploited the topological selection criteria applied in the $\Dzero$ meson analysis.
The $\Dstar$ signal was observed calculating the invariant-mass difference $\Delta m=m_{\rm D^{*+}}-m_{\rm D^0}$ between the reconstructed $\Dstar$ and the decay $\Dzero$, as a narrow peak at $\Delta m \approx 145.4~\mev/c^2$ close to the threshold and thus in a rather low combinatorial background region. Furthermore, the resolution in $\Delta m$ is mostly defined by the pion momentum resolution.  

The procedure for the track reconstruction in the ALICE central detectors is explained in Ref.~\cite{aliceJINST}. The details concerning D meson decay tracks reconstruction are described in Ref.~\cite{aliceDmeson7TeV}. 
Secondary vertices of $\Dzero$ and $\Dplus$ meson candidates were reconstructed, with the same algorithm used to compute the primary vertex, from tracks having $|\eta| < 0.8$, $\pt > 0.3$~$\gevc$. Tracks were also required to have at least 70 space points (out of a maximum of 159) and $\chi^2/{\rm ndf} < 2$ in the TPC, and at least one hit in either of the two layers of the SPD. 
Only tracks compatible with a kaon or a pion were kept. The particle identification criteria consisted in a $3\sigma$ compatibility cut between the measured and expected signals, using the specific energy deposit and the time-of-flight from the TPC and TOF detectors, respectively. This conservative strategy was aimed at keeping $\sim100\%$ of the signal (see Sec.~\ref{sec:systematics}). 
Exception was done for $\Dzero$ with $\pt<2~\gev/c$, where stricter requirements for the decay kaon identification in either the TPC or the TOF detectors were considered. These requirements were dependent on the track momentum. 
Tracks with no associated signal in the TOF detector were identified using only the TPC information. Tracks with contradictory responses from the TPC and TOF detectors were considered as unidentified and included in the analysis as compatible with both a pion and a kaon. 
The pions from the $\Dstar$ candidate decay were required to have a minimum transverse momentum of $100$~MeV/$c$ and a minimum of 3 (out of 4) associated clusters in the ITS, in addition to the TPC quality criteria mentioned above. Particle identification was not applied to pion tracks from the $\Dstar$ candidate decay. 

The topological selection criteria considered for the three mesons are described in the following. The selection values are $\pt$ dependent and were adjusted to optimize the statistical significance of the signal, while keeping the selection efficiency increasing with $\pt$. For illustration, the selection values applied for $\Dzero$ and $\Dplus$ mesons at low $\pt$, and typical values for the $\Dstar$ selection, are explained in the next paragraphs. 

$\Dzero$ mesons were reconstructed from combinations of two tracks with a minimum transverse momentum of $0.4~\gevc$. The two tracks impact parameter (distance of closest approach of the track to the primary interaction vertex, $d_0$) significance in the bending plane ($r\varphi$) was of $|d_0|/\sigma_{d_{0}} > 0.5$, and the two track distance of their closest approach was smaller than $300~\mum$. 
Only the candidates associated with secondary vertices with a minimum displacement of $100~\mum$ from the primary vertex were retained. 
In addition, the cosine of the angle ($\theta^{*}$) between the kaon momentum in the $\Dzero$ rest frame and the $\Dzero$ boost direction was required to be $|\cos \theta^{*}|<0.8$, the product of the $\Dzero$ decay track impact parameters was set to $d_{0}^{\pi}\times d_{0}^{K}~<~-(250~\mum)^2$, 
and the angle ($\theta_{\rm pointing}$) between the $\Dzero$ reconstructed momentum and its flight line (vector between the primary and secondary vertices) was constrained by $\cos \theta_{\rm pointing} > 0.8$. 

The $\Dplus$ meson topological selection was similar to the one of the $\Dzero$. This being a three body decay, a looser cut on the $\pt$ of the decay tracks of 0.3~$\gevc$ was applied. The topological selection of the candidates was tighter than for the $\Dzero$ in order to deal with the large combinatorial background. 
The main selection variables were: decay length larger than 800$~\mum$, $\cos \theta_{\rm pointing} >0.92$ and the sum of the square of the decay track impact parameters  $\Sigma~d_{0}^{2} >  (500 \mum)^2$. 

The $\Dstar$ candidates were reconstructed by applying kinematical selections on the final decay products and on the topology of the $\Dzero$ decay. The $\Dzero$ decay candidates were selected from pairs of tracks with similar criteria to that applied for the $\Dzero$ analysis and described above. The selection values vary in the $\Dstar$ candidates $\pt$ interval. 
In particular, the angle between the $\Dzero$ reconstructed momentum and its flight line was kept to $\cos \theta_{\rm pointing} > 0.9$ for candidates with $\Dstar$ $\pt < 4$~$\gevc$ while it was released to $0.7$ for higher $\pt$ profiting from the low combinatorial background. A typical value of the product of the track impact parameters was $d_{0}^{\pi}\times d_{0}^{K}~<~-(60~\mum)^2$.

\begin{figure}[!htbp]  
\begin{center}        
\includegraphics[width=1.0\textwidth]{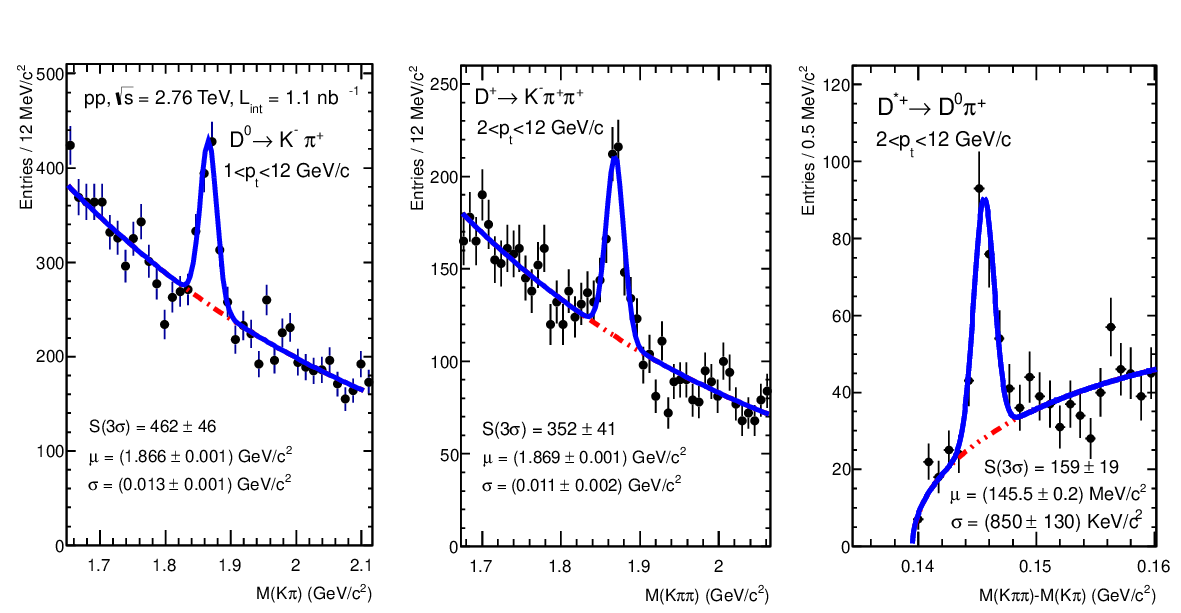}
\caption{Invariant-mass spectrum of $\Dzero+\Dzerobar$~(left) and $\Dplus+\Dminus$~(centre) candidates, and invariant-mass difference, $\Delta m = m_{K \pi \pi} - m_{K \pi}$, for $\Dstar+\Dstarm$ candidates~(right) in pp collisions at $\sqrts=2.76~$TeV.}
\label{fig:invmass}
\end{center}
\end{figure}

After the $\Dzero$, $\Dplus$, and $\Dstar$ candidates were reconstructed with the above kinematical and topological cuts and particle identification criteria, a fiducial acceptance cut $|y_{\rm D}|<y_{\rm fid}{\rm(}\pt\rm{)}$ was applied, with $y_{\rm fid}$ increasing with a polynomial form from 0.5 to 0.8 in $0 < \pt < 5$~$\gevc$ and $y_{\rm fid}= 0.8$ above 5~$\gevc$. 
The $\Dzero$, $\Dplus$, and $\Dstar$ raw yields for particle plus anti-particle are summarized in Table~\ref{tab:Yields} for each $\pt$ interval. 
They were obtained by fitting the invariant-mass distribution with a Gaussian distribution to describe the signal and an ad-hoc function for the background (see Fig.~\ref{fig:invmass}). For the $\Dzero$ and $\Dplus$, the background was reproduced by an exponential function, while for $\Dstar$ the convolution of the exponential with a threshold function was used. 
The D meson peak width was measured to be 10--20~MeV$/c^2$ for $\Dzero$ and $\Dplus$, and 600--900~KeV$/c^2$ for $\Dstar$ mesons, increasing with transverse momentum. These widths reflect the 1--2\% momentum resolution for the decay tracks in the relevant $\pt$ range.

\begin{table}[!htbp]
\caption{Measured $\Dzero$, $\Dplus$, and $\Dstar$ raw counts and their charge conjugates in \pp~collisions at $\sqrts=2.76$~TeV per $\pt$ interval with an integrated luminosity $\mathcal{L}_{\rm int}=1.1$~$\nbinv$. The systematic uncertainty estimate is described in section~\ref{sec:systematics}.} 
\centering
\begin{tabular}{cccc} 
\hline \hline
$\pt$ interval
 & \multicolumn{3}{c}{$N^{\rm raw}$ $\pm$ stat. $\pm$ syst.} \\
 ($\gevc$) 
 & $\Dzero+\Dzerobar$ & $\Dplus+\Dminus$ & $\Dstar+\Dstarm$\\
\hline 
1--2 &   \phantom~48 $\pm$ 18 $\pm$ \phantom~7 & --  & -- \\
2--4 &   201 $\pm$ 32 $\pm$ 30 & \phantom~98  $\pm$ 24 $\pm$ 10 &  53 $\pm $15 $\pm$ 7 \\
4--6 &  116 $\pm$ 19 $\pm$ 17 &   123 $\pm$ 23 $\pm$ 12 &  50 $\pm$ \phantom~9 $\pm$ 6 \\
6--8 &   \phantom~74 $\pm$ 19 $\pm$ 11 &  \phantom~62 $\pm$ 16  $\pm$ 6 &  30 $\pm$ \phantom~6 $\pm$ 2 \\ 
8--12 & \phantom~38 $\pm$ 11 $\pm$ \phantom~6 &  \phantom~30 $\pm$ \phantom~9 $\pm$ 5 &  23 $\pm$ \phantom~7 $\pm$ 2 \\
\hline \hline
\end{tabular}
\label{tab:Yields}
\end{table}

\subsection{Corrections and systematic uncertainties}
\label{sec:corrections}
\begin{figure}[!htbp]  
\begin{center}        
\includegraphics[width=\textwidth]{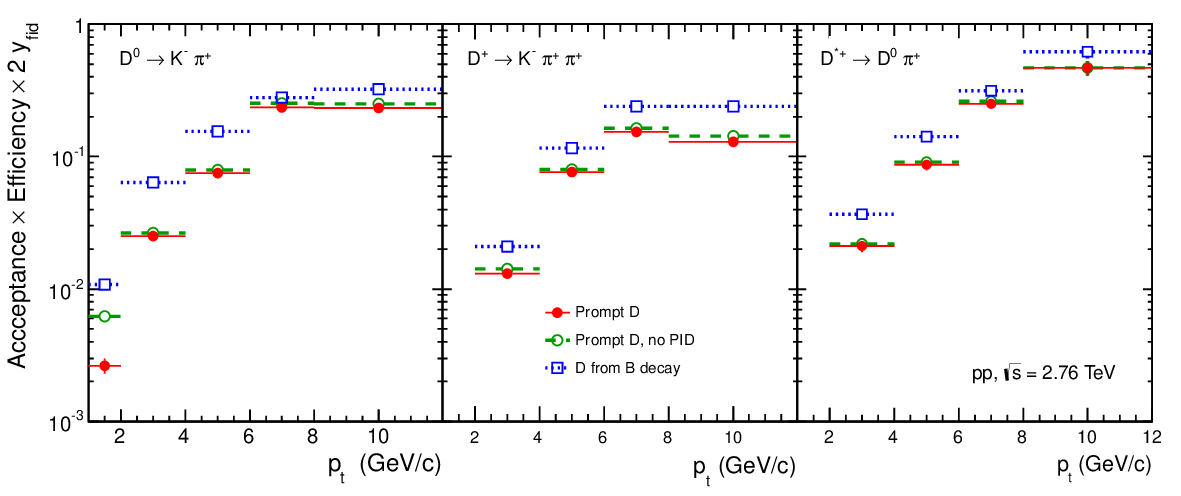}
\caption{Acceptance $\times$ efficiency $\times 2 y_{\rm fid}$ for $\Dzero$~(left), $\Dplus$~(centre) and $\Dstar$~(left) as a function of $\pt$, where $2 y_{\rm fid}$ is the fiducial acceptance (see text).}
\label{fig:eff}
\end{center}
\end{figure}

The production cross sections of prompt charmed mesons were calculated as (e.g. for $\Dplus$):
\begin{equation}
  \label{eq:crosssectionD}
  \left.\frac{{\rm d} \sigma^{\rm D^+}}{{\rm d}\pt}\right|_{|y|<0.5}=
  \frac{1}{2}\frac{1}{\Delta y\,\Delta\pt}\frac{\left.f_{\rm prompt}(\pt)\cdot N^{\rm D^\pm~raw}(\pt)\right|_{|y|<y_{\rm fid}(\pt)}}{({\rm Acc}\times\epsilon)_{\rm prompt}(\pt) \cdot{\rm BR} \cdot \mathcal{L}_{\rm int}}\,.
\end{equation}
The raw yields, $N^{\rm D^\pm~raw}(\pt)$, listed in Table~\ref{tab:Yields}, were corrected for the B feed-down contribution, $f_{\rm prompt}(\pt)$, the fiducial acceptance, $\Delta y= 2 \, y_{\rm fid}$, and the experimental acceptance and reconstruction efficiency, $({\rm Acc}\times\epsilon)_{\rm prompt}(\pt)$~shown in Fig.~\ref{fig:eff}. 
The rapidity acceptance correction, using the factor $2 \, y_{\rm fid}$, assumes that the rapidity distribution of D mesons is uniform in the range $|y| < y_{\rm fid}$. This assumption was verified using the PYTHIA 6.4.21~\cite{pythia} event generator with Perugia-0 tune~\cite{perugia0} and the FONLL pQCD calculation~\cite{fonll,FONLLalice}. Both calculations generate a D meson yield that is uniform within 1\% in the range $|y| < 0.8$. 
The efficiencies were calculated from a Monte Carlo simulation using the PYTHIA 6.4.21~\cite{pythia} event generator with the Perugia-0 tune~\cite{perugia0} and GEANT3~\cite{geant3}. The LHC beam conditions and the apparatus configuration (inactive channels, noise, calibration, alignment) were considered taking into account their evolution with time. 
The acceptance correction was evaluated to account for the fiducial rapidity cut, $y_{\rm fid}$. 
A fraction of the reconstructed D mesons comes from B meson decays. Since these are characterized by a relatively long life time (B meson c$\tau \approx$~460--490~$\mu$m~\cite{pdg}), the decay tracks of D mesons from B decays are further displaced from the primary vertex, and the selection criteria enhance their relative contribution to the raw yields. 
Their contribution was evaluated using FONLL pQCD calculations~\cite{fonll,FONLLalice} of the beauty production cross section and the B~$\rightarrow$~D decay kinematics from the EvtGen package~\cite{evtgen}, with the procedure described in Ref.~\cite{aliceDmeson7TeV}, and was subtracted from the raw yields. The prompt fraction, $f_{\rm prompt}$, ranges within $88\%$ to $98 \%$ depending on the D meson and $\pt$ interval. 
These FONLL pQCD calculations were chosen since they reproduce the Tevatron~\cite{fonllBcdf} and LHC~\cite{lhcbBeauty,cmsJpsi} measurements discussed above.
The corrected yields were then divided by a factor of two to obtain the averaged yield of the D mesons and their charge conjugates. 
They were finally normalized by their decay branching ratio, BR, and the sample integrated luminosity $\mathcal{L}_{\rm int} = 1.1$~nb$^{-1}$~\cite{oyama}. 

\label{sec:systematics}
The systematic uncertainties are summarized in Table~\ref{tab:Syst} for the lowest and highest $\pt$ interval for each meson species. 
The systematic uncertainties from the yield extraction were determined, in each $\pt$ interval, by studying the yield variation as obtained with a different background fit function (exponential, polynomial, linear) and by counting the signal in the candidates invariant mass range ($\pm 3 \sigma$) after subtracting the background (evaluated by fitting the distribution side bands). The yield extraction systematic uncertainties were set to one half of the full spread of the yield variation in each $\pt$ interval. 
The uncertainty on the single track efficiency was evaluated to be 5\% (per track), by taking into account the influence of the track finding in the TPC, the prolongation from the TPC to ITS, and the track quality criteria. 
This was estimated by comparing the relative variation of the efficiency in data and simulations and by varying the track selection criteria.  
This track efficiency uncertainty results into an uncertainty of 10\% for the two-body decay of $\Dzero$ mesons and of 15\% for the three-body decay of $\Dplus$ and $\Dstar$ mesons.
The influence of the analysis cuts amounts to about 15\%, and accounts for possible discrepancies of the D meson selection variables in data and simulations. The distributions of these variables in data, dominated by background candidates, and in simulations were compared and found to be compatible. 
The uncertainty was estimated by repeating the analysis with different sets of cuts and set to one half of the spread of the corrected yields. 
The uncertainty associated to the particle identification (PID) selections was studied by comparing the corrected invariant yields with and without PID. 
As the $\pt$ bins used in this analysis are finite and the transverse momentum distribution of the candidates is steep, the Monte Carlo $\pt$ shape could influence the acceptance and efficiency corrections per $\pt$ bin. These corrections were computed with different $\pt$ shapes (PYTHIA, FONLL, flat $\pt$) and their influence was found to be 3\% for D meson with $\pt<2$~$\gevc$, and of 1\% for larger $\pt$. 
The particle and anti-particle yields were evaluated independently per $\pt$ bin and found to be in agreement within statistical uncertainties (of about 30\% for the $\Dzero$ analysis). 
The systematic uncertainty on the subtraction of the B feed-down contribution, explained in detail in Ref.~\cite{aliceDmeson7TeV}, accounts for the full variation of this correction considering either only the FONLL B feed-down prediction or the ratio of the prompt and B feed-down calculations, and includes the uncertainties of the theoretical calculations. 
Finally, the uncertainties of the correction for the D meson decay branching ratio (BR) and the luminosity (overall normalization) were evaluated as described above, and are presented in Table~\ref{tab:Syst}.

\begin{table}[!htbp]
\caption{Summary of relative systematic uncertainties for given $\pt$ intervals for each meson species.} 
\centering
\begin{tabular}{lccccccc} 
\hline \hline
 & \multicolumn{3}{c}{$\Dzero$} 
 & \multicolumn{2}{c}{$\Dplus$} 
 & \multicolumn{2}{c}{$\Dstar$}\\
 $\pt$ interval (GeV/$c$)
 & 1--2  & 2--4 & 8--12
 & 2--4 &  8--12 
 & 2--4 &  8--12 \\
\hline 
Yield extraction      & 15\%  & 15\% & 15\%
                      & 10\%  & 15\% 
                      & 14\%  & 6\% \\
Tracking efficiency   & \multicolumn{3}{c}{10\%} 
                      & \multicolumn{2}{c}{15\%}
                      &  \multicolumn{2}{c}{15\%}\\
Cut efficiency        & 20\%  & 10\% & 15\% 
                      &  15\% &  15\% 
                      &  15\% &  10\% \\
PID efficiency        & 15\% & 15\% & \phantom~5\% 
                      & 10\% &  \phantom~5\% 
                      &  \phantom~5\% & \phantom~5\% \\
MC $\pt$ shape       & 3\% &  1\% & 1\%
			 & 1\% & 1\%
			  & 1\% & 1\% \\
Feed-down from B      & $^{\phantom~+3}_{-39}$\%  & $^{\phantom~+3}_{-17}$\%  & $^{+3}_{-7}$\% 
                      & $^{\phantom~+2}_{-12}$\% & $^{\phantom~+4}_{-13}$\% 
                      & $^{\phantom~+2}_{-10}$\% &  $^{+3}_{-5}$\% \\
Branching ratio       & \multicolumn{3}{c}{1.3\%} 
                      & \multicolumn{2}{c}{2.1\%} 
                      & \multicolumn{2}{c}{1.5\%}\\
Normalization         & \multicolumn{3}{c}{1.9\%} 
		               & \multicolumn{2}{c}{1.9\%}
			      & \multicolumn{2}{c}{1.9\%} \\
\hline \hline
\end{tabular}
\label{tab:Syst}
\end{table}

\section{D meson cross section at $\sqrts=$ 2.76~TeV}
\label{sec:crosssection}
\begin{figure}[!htbp]  
\begin{center}        
\includegraphics[width=\textwidth]{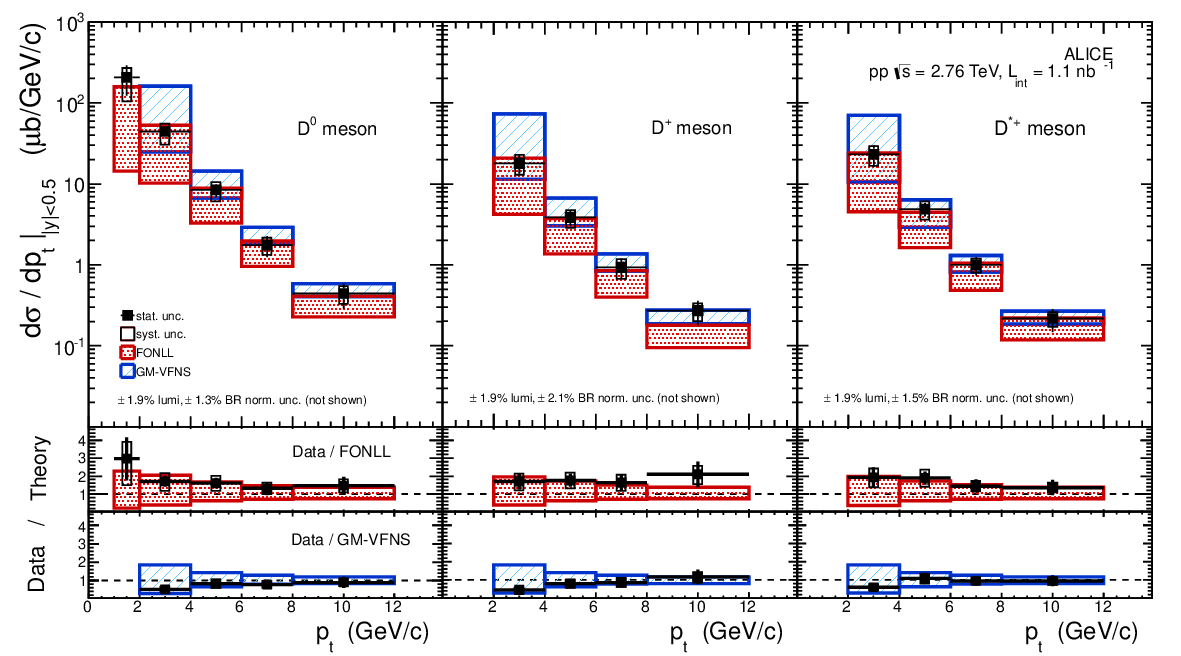}
\caption{Top : $\pt$-differential cross section for prompt $\Dzero$, $\Dplus$, and 
$\Dstar$ mesons in pp collisions at $\sqrts=2.76~$TeV compared with FONLL~\cite{fonll,FONLLalice} 
and GM-VFNS~\cite{gmvfns,VFNSalice} theoretical predictions. Bottom: the ratio of the measured cross section and the central FONLL and GM-VFNS calculations.}
\label{fig:ptdiffcs}
\end{center}
\end{figure}

The prompt $\Dzero$, $\Dplus$, and $\Dstar$ mesons $\pt$-differential cross sections were derived from the raw yields as described in section~\ref{sec:corrections}. 
The global systematic uncertainties were evaluated summing in quadrature the various uncertainty sources explained in section~\ref{sec:systematics}, and reported in Table~\ref{tab:Syst}. 
The results are summarized in Table~\ref{tab:ptdiffcs} and shown in Fig.~\ref{fig:ptdiffcs}. The top panels of the figures present the measurement together with the FONLL~\cite{fonll,FONLLalice} and the GM-VFNS~\cite{gmvfns,VFNSalice} theoretical predictions, while the bottom panels represent the ratio of the measured cross section and the calculations. 
Both calculations use the CTEQ~6.6 parton distribution functions (PDFs)~\cite{cteq6.6}, and vary the factorization and renormalization scales, $\mu_{\rm F}$ and $\mu_{\rm R}$, around their central values of $\muf = \mur = m_{\rm t}$ in the ranges $0.5 < \muf/m_{\rm t} < 2$, $0.5 < \mur/m_{\rm t} < 2$, with the constraint $0.5 < \muf/\mur < 2$, where $m_{\rm t} = \sqrt{ \pt^2 +\mc^2}$. 
The FONLL calculation varies the charm quark mass within $1.3 < \mc < 1.7~\gevc^2$ while GM-VFNS assumes $\mc = 1.5~\gevc^2$. 
The FONLL and GM-VFNS theoretical predictions are compatible with the measurements within the experimental and theoretical uncertainties. 
Nevertheless, it can be noted that the central prediction of FONLL tends to underestimate charm production whereas the central GM-VFNS calculation seems to overestimate it, as seen in the lower panels of Fig.~\ref{fig:ptdiffcs}. 
This behaviour is in accordance with our results on the prompt $\Dzero$, $\Dplus$, and $\Dstar$ mesons $\pt$-differential cross sections at $\sqrts=7$~TeV~\cite{aliceDmeson7TeV}.

The  visible cross sections of prompt D mesons ($\sigma^{\rm D}_{\rm vis}$), i.e. the $\pt$-integrated production cross sections in the rapidity range $|y|<0.5$, where the measurement was performed, are reported in Table~\ref{tab:D-cross-section-visible} for both the present result at $\sqrts=2.76$~TeV and for $\sqrts=7$~TeV~\cite{aliceDmeson7TeV}. 
\begin{table}[!htbp]
\caption{
\label{tab:ptdiffcs}
Production cross section in $|y|<0.5$ for prompt $\Dzero$, $\Dplus$, and 
$\Dstar$ mesons in pp collisions at $\sqrts=2.76~$TeV, in transverse momentum intervals.
The branching ratio uncertainty and the uncertainty in the normalization of $1.9\%$ are not included in the systematic uncertainties
reported in this table.}
\centering
\begin{tabular}{cccc} 
\hline \hline
$\pt$ interval ($\gevc$) & \multicolumn{3}{c}{d$\sigma$/d$\pt~\pm$ stat. $\pm$ syst. ($\mu$b / $\gevc$)} \\
 & $\Dzero$ & $\Dplus$ & $\Dstar$\\
\hline 
1--2 &   207 $\pm$ 84 $^{+64}_{-103}$ &--& --\\
2--4 &   44.1 $\pm$ 7.7 $^{+11}_{-14}$ 	&  18.0 $\pm$ 4.6 $^{+4.6}_{-5.1}$ &   23.2 $\pm$ 6.9 $^{+6.0}_{-6.5}$ \\[0.5ex]
4--6 &   8.4 $\pm$ 1.5 $^{+2.2}_{-2.3}$ 	&  3.82 $\pm$ 0.77 $^{+0.92}_{-0.97}$ &   4.90 $\pm$ 0.95 $^{+1.22}_{-1.26}$ \\[0.5ex]
6--8 &   1.75 $\pm$ 0.50 $^{+0.42}_{-0.43}$	 &   0.93 $\pm$ 0.26 $^{+0.25}_{-0.26}$ &   1.00 $\pm$ 0.26 $^{+0.20}_{-0.20}$ \\[0.5ex]
8--12 &  0.44 $\pm$ 0.15 $^{+0.11}_{-0.11}$ &   0.27 $\pm$ 0.09 $^{+0.06}_{-0.07}$ &  0.22$\pm$ 0.07 $^{+0.04}_{-0.04}$ \\[0.5ex]
\hline \hline
\end{tabular}
\end{table}

\begin{figure}[!htbp]  
\begin{center}        
\includegraphics[width=\textwidth]{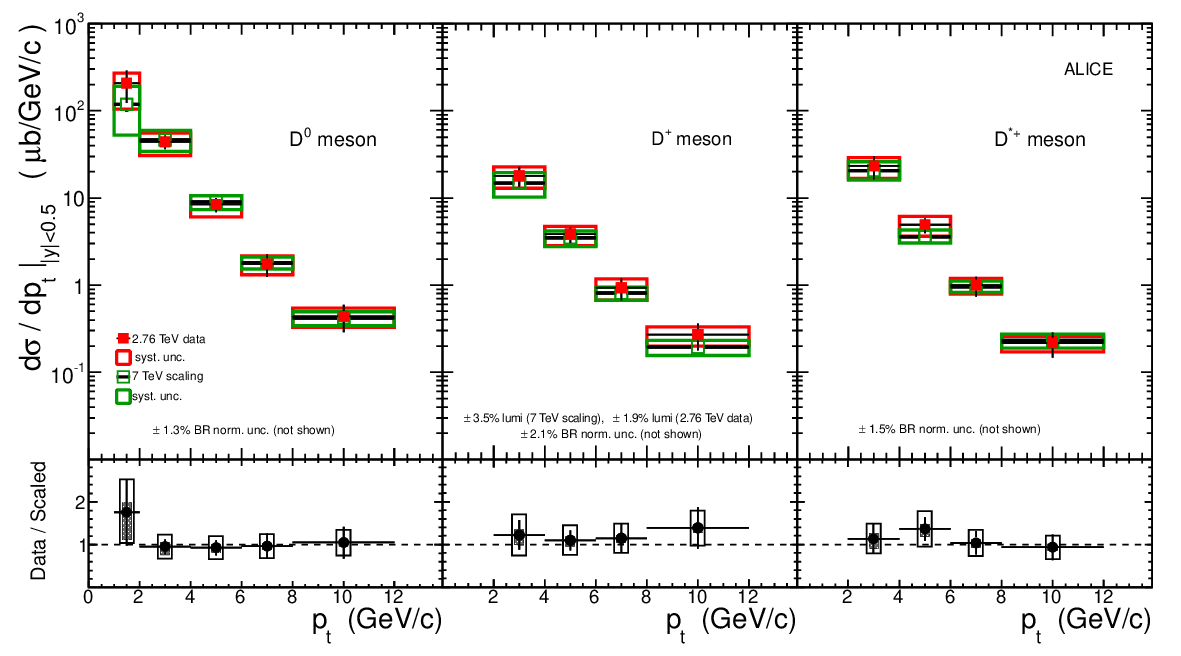}
\caption{Top : $\pt$-differential cross section for prompt $\Dzero$, $\Dplus$, and 
$\Dstar$ mesons in pp collisions at $\sqrts=2.76~$TeV compared with the scaling of the ALICE measurement at $\sqrts=7~$TeV.
Bottom : Ratio of the $\sqrts=2.76~$TeV cross section and the $\sqrts=7~$TeV measurement scaling, where the filled boxes represent the scaling uncertainties and the empty boxes the measurement systematics.}
\label{fig:ptdiffcsExt}
\end{center}
\end{figure}

The measurements in pp collisions $\sqrts=2.76$~TeV described here provide a baseline for the studies of the QCD matter created in \PbPb~collisions at the same centre-of-mass energy~\cite{aliceDmesonRaapaper}. 
However, since the statistics is limited and does not allow a comparison with the \PbPb~measurements for every $\pt$~interval, the reference used for comparisons of the \PbPb~and~\pp~yields was obtained from a pQCD--based (FONLL) energy scaling of the 7~TeV $\pt$-differential cross sections to 2.76~TeV~\cite{aliceDmeson7TeV,EnergyExtrap}.
The scaling factor was evaluated from the ratio of the theoretical cross sections at these energies and the uncertainties were determined by the envelope of the scaling factors obtained by varying the calculation parameters ($\mc$, $\muf$ and $\mur$) as described above. 
The 2.76~TeV $\pt$-differential cross sections and the results of the energy scaling are shown in Fig.~\ref{fig:ptdiffcsExt}, where the 7~TeV measurements were rebinned to match the 2.76~TeV $\pt$-binning. 
The agreement is remarkable in all $\pt$ bins. The results are compatible within statistical uncertainties only, and their central values coincide within $5$--$10\%$ in almost all $\pt$ bins, confirming the stability and appropriateness of the energy scaling procedure.  

\section{Total charm cross section}
\label{sec:charmAndPv}
The measured cross sections were extrapolated to the full phase space by scaling the measured cross section by the ratio of the total cross section over the cross section in the experimentally covered phase space calculated with the FONLL central parameters. 
Here the visible cross sections for $\pt>1$~$\gevc$ and $|y|<0.5$ were considered for \pp~collisions at 7 and 2.76~TeV (see Table~\ref{tab:D-cross-section-visible}). 
Systematic uncertainties of the calculation were estimated varying the renormalization ($\mur$) and factorization ($\muf$) scale variables, and the charm quark mass ($\mc$) as described in the previous section. 
Uncertainties in the parton distribution functions were estimated using the CTEQ6.6~\cite{cteq6.6} PDF uncertainties eigenvectors  and adding the largest positive and negative variation in quadrature. 

The total charm production cross section was estimated for each species of D meson separately by dividing the total D meson production cross section by the relative production yield for a charm quark hadronizing to a particular species of D meson, that is the fragmentation fractions (FF) of $0.557 \pm 0.023$ for $\Dzero$, $0.226 \pm 0.010$ for $\Dplus$, and $0.238 \pm 0.007$ for $\Dstar$~\cite{pdg}. 
The measured yields are consistent with these ratios, as can be seen in Table~\ref{tab:D-cross-section-pt}. 
We then calculated the weighted average of the total charm production cross section from the extrapolated values for  $\Dzero$, $\Dplus$, and  $\Dstar$.

\begin{table}[!htbp]
\caption{Visible production cross sections of prompt D mesons, $\sigma^{\rm D}_{\rm vis}(|y|<0.5)$ in pp collisions at $\sqrts=2.76$ and $7$~TeV. 
The normalization systematic uncertainty of 1.9\%~(3.5\%) at $\sqrts=2.76$~(7)~TeV and the decay BR uncertainties are not quoted here.}
\centering
\begin{tabular}{lcccc} 
\hline\hline 
Meson & $\sqrt{s}\ (\tev)$ & $\pt$ interval ($\gevc$)
	& $\sigma^{\rm D}_{\rm vis} \pm$ stat. $\pm$ syst. ($\mub$)  \\ \hline
$\Dzero$	&	2.76 & 	1--12 &  317 $\pm$ 85 $^{\phantom~+72}_{-120}$ \\[0.5ex]
$\Dplus$ 	& 	2.76 & 	2--12 & \phantom~47 $\pm$ \phantom~\phantom~9 $^{\phantom~+10}_{\phantom~-12}$  \\[0.5ex]
$\Dstar$	&	2.76 & 	2--12 &  \phantom~59 $\pm$ 14 $^{\phantom~+13}_{\phantom~-14}$  \\[0.5ex]
$\Dzero$	&	7 & 	1--16 &  412 $\pm$ 33 $^{\phantom~+55}_{-140}$ \\[0.5ex]
$\Dplus$ 	& 	7 & 	1--24 & 198 $\pm$ 24 $^{\phantom~+42}_{\phantom~-73}$ \\[0.5ex]
$\Dstar$	&	7 & 	1--24 & 203 $\pm$ 23 $^{\phantom~+30}_{\phantom~-67}$\\[0.5ex]
\hline\hline
\end{tabular}
\label{tab:D-cross-section-visible}
\end{table}
\begin{table}[!htbp]
\caption{Production cross sections ${\rm d} \sigma^{\rm D} / {\rm d }y$ ($\mub$) of D mesons, integrated over all $\pt$ for $|y|<0.5$.}
\centering
\begin{tabular}{lcccccccc} 
\hline\hline 
Meson & $\sqrt{s}\ (\tev)$ & ${\rm d} \sigma^{\rm D} / {\rm d} y$ & stat.  & syst. & lum. & BR & extr.  \\
\hline 
$\Dzero$  & 2.76 & 428 & $\pm 115$ & $^{+98}_{-163}$ & $\pm 8$ & $\pm 6$   & $^{+151}_{-20}$ \\[0.5ex]
$\Dplus$  & 2.76 &  127 & $\pm 26$ & $^{+28}_{-31}$ & $\pm 2$ & $\pm 3$   & $^{+38}_{-23}$ \\[0.5ex]
$\Dstar$ 	& 2.76 & 148 & $\pm 35$ & $^{+33}_{-36}$ & $\pm 3$ & $\pm 2$  & $^{+42}_{-23}$\\[0.5ex]
$\Dzero$  & 7 & 516 & $\pm 41$ & $^{+69}_{-175}$ & $\pm 18$ & $\pm 7$  & $^{+120}_{-37}$ \\[0.5ex]
$\Dplus$  & 7 & 248 & $\pm 30$ & $^{+52}_{-92}$ & $\pm 9$ & $\pm 5$  & $^{+57}_{-18}$ \\[0.5ex]
$\Dstar$   & 7 & 247 & $\pm 27$ & $^{+36}_{-81}$ & $\pm 9$ & $\pm 4$  & $^{+57}_{-16}$ \\[0.5ex]
\hline
\end{tabular}
\label{tab:D-cross-section-pt}
\end{table}
\begin{table}[!htbp]
\caption{Total production cross sections $\sigmatot^{\rm D} (\mb)$ of D mesons, extrapolated to the full phase space.}
\centering
\begin{tabular}{lccccccccc}
\hline\hline 
Meson & $\sqrt{s}\ (\tev)$ & $\sigmatot^{\rm D}$ & stat. & syst.  & lum.  & BR & extr.\\ 
\hline 
$\Dzero$     & 2.76 & 3.13 & $\pm 0.84$ & $^{+0.71}_{-1.19}$ & $\pm 0.06$ & $\pm 0.04$ &  $^{+2.02}_{-0.14}$ \\[0.5ex]
$\Dplus$     & 2.76 &  0.93 & $\pm 0.19$ & $^{+0.20}_{-0.22}$ & $\pm 0.02$ & $\pm 0.02$ &  $^{+0.41}_{-0.09}$ \\[0.5ex]
$\Dstar$ & 2.76 & 1.08 & $\pm 0.25$ & $^{+0.24}_{-0.26}$ & $\pm 0.02$ & $\pm 0.02$ &  $^{+0.51}_{-0.10}$\\[0.5ex]
$\Dzero$ & 7 & 4.42 & $\pm$ 0.35 & $^{+0.59}_{-1.50}$ & $\pm$ 0.15 & $\pm$ 0.06 &  $^{+2.59}_{-0.19}$ \\[0.5ex]
$\Dplus$ & 7 & 2.12 & $\pm$ 0.26 & $^{+0.45}_{-0.78}$ & $\pm$ 0.07 & $\pm$ 0.04 &  $^{+1.23}_{-0.09}$ \\[0.5ex]
$\Dstar$ & 7 & 2.11 & $\pm$ 0.24 & $^{+0.31}_{-0.70}$ & $\pm$ 0.07 & $\pm$ 0.03 &  $^{+1.24}_{-0.08}$ \\[0.5ex]
\hline\hline
\end{tabular}
\label{tab:D-cross-section-full}
\end{table}

The extrapolated cross sections for D mesons are given in Tables~\ref{tab:D-cross-section-pt}~and~\ref{tab:D-cross-section-full}  with uncertainties resulting from the yield extraction (stat.),  the quadratic sum of the experimental uncertainty in D-meson reconstruction and the uncertainty in subtracting the contribution of D mesons originating from beauty production (syst.), a 1.9\% (3.5\%) uncertainty in the absolute luminosity (lum.) at 2.76~TeV (7~TeV), the uncertainty in the branching ratios (BR), and the uncertainty due to extrapolation to the full phase space (extr.).

The ratio, $\Pv$, of $\cdbar$ D mesons produced in a vector state to those produced in a vector or a pseudoscalar state was calculated by taking the ratio of $\sigmatot^{\Dstar}$ to the sum of $\sigmatot^{\Dstar}$ and the part of  $\sigmatot^{\Dplus}$ not originating from $\Dstar$ decays,
 \begin{equation}
\Pv
= \frac{\sigmatot(\Dstar)}{\sigmatot (\Dstar)+\sigmatot (\Dplus)
- \sigmatot(\Dstar) \cdot (1 - {\rm BR}_{\Dstar\rightarrow {\rm D}^0\pi^+})}
= \frac{\sigmatot(\Dstar)}{\sigmatot(\Dplus)+\sigmatot (\Dstar)\cdot {\rm BR}_{\Dstar\rightarrow {\rm D}^0\pi^+}}.
\label{eq:extrap_pv}
\end{equation}

\begin{figure}[!htbp]
\begin{center}
\includegraphics[width=0.495\textwidth,height=0.45\textwidth]{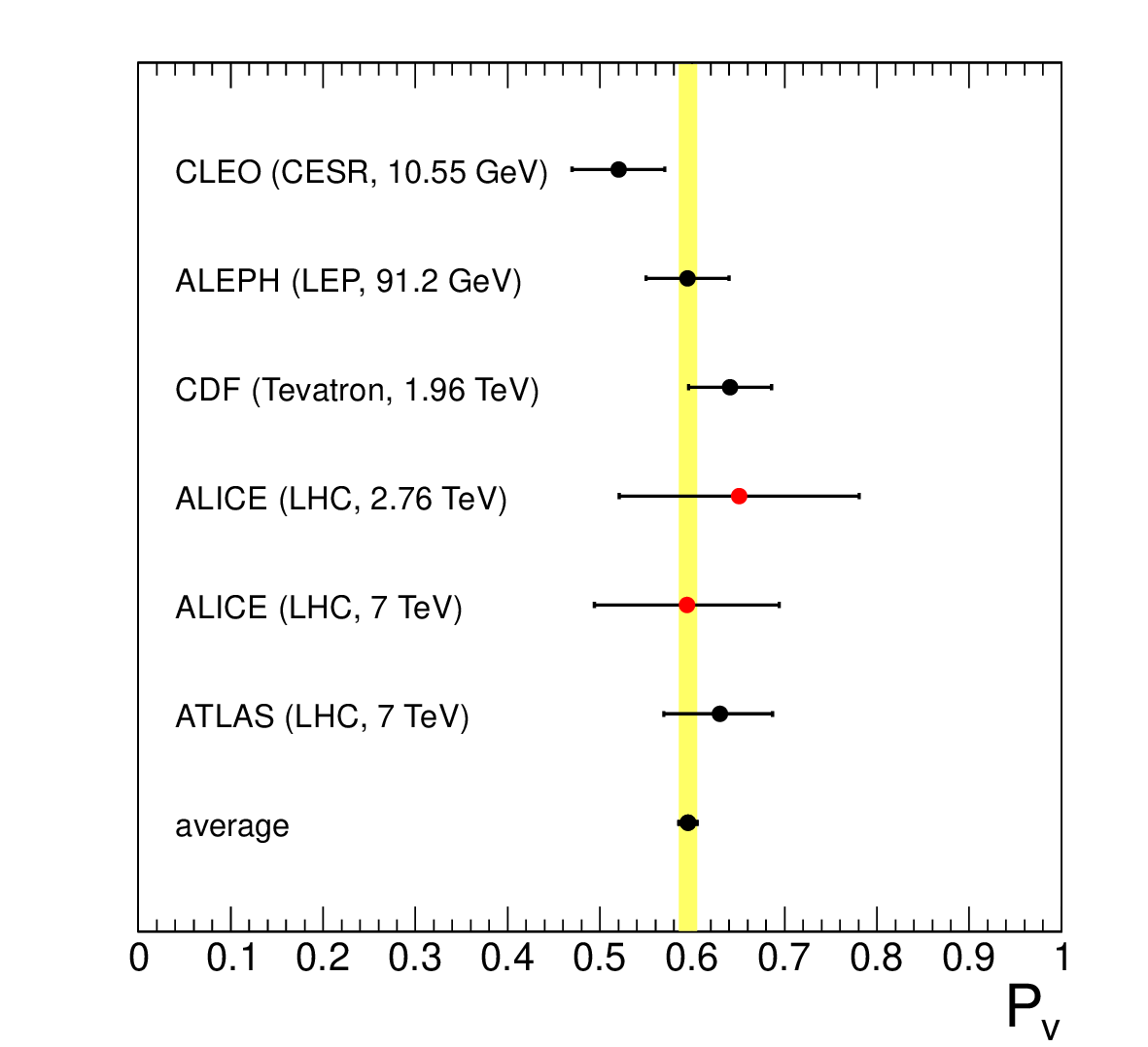} 
\includegraphics[width=0.495\textwidth,height=0.45\textwidth]{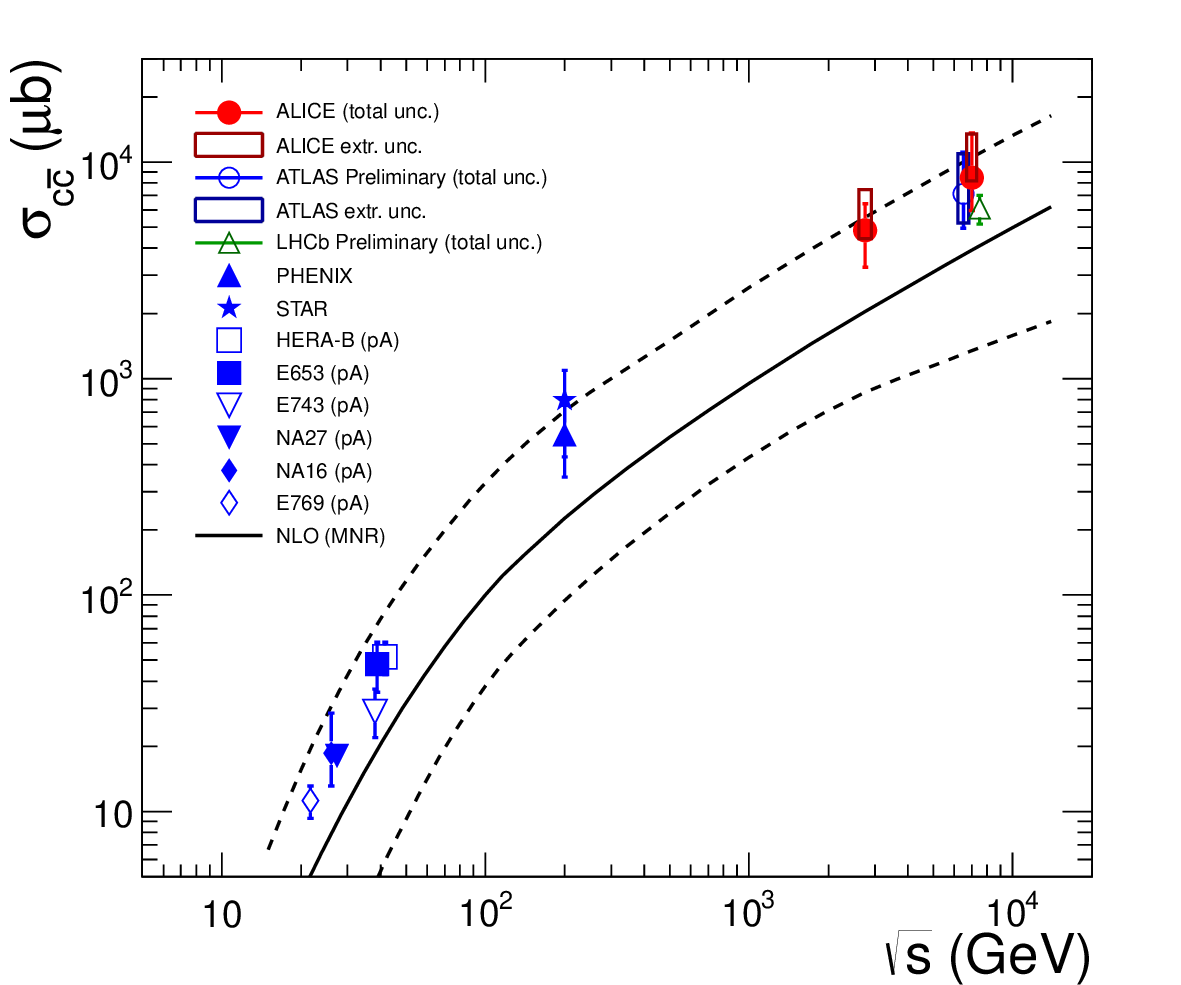}
\caption{
\label{fig:pv}
Left: The fraction $\Pv$ of $\cdbar$ D mesons created in a vector state to vector and pseudoscalar prompt D mesons~\cite{CLEO,ALEPH,charmcdf,CDFPv,STARDmeson,aliceDmeson7TeV,ATLASDmeson}. The weighted average of the experimental measurements reported in Ref.~\cite{ADavid} and of the LHC data~\cite{aliceDmeson7TeV,ATLASDmeson} shown in the figure is $\Pv = 0.60 \pm 0.01$, and is represented by a solid yellow vertical band. 
\label{fig:enercanvfull}
Right: Energy dependence of the total nucleon--nucleon charm production cross section~\cite{Lourenco,STAR,ATLASDmeson,LHCbDmeson,PHENIX}. In case of proton--nucleus (pA) or deuteron--nucleus (dA) collisions, the measured cross sections have been scaled down by the number of binary nucleon--nucleon collisions  calculated in a Glauber model of the proton--nucleus or deuteron--nucleus collision geometry. The NLO MNR calculation~\cite{MNR}~(and its uncertainties) is represented by solid (dashed) lines.
} 
\end{center}
\end{figure}

The obtained values are: 
\begin{eqnarray}
\Pv (\mathrm{2.76~TeV}) = 0.65 \pm 0.10 \, (\mathrm{stat.}) \,  \pm 0.08 \, (\mathrm{syst.} ) \, \pm 0.010 \, (\mathrm{BR}) \, ^{+0.011}_{-0.004} (\mathrm{extr.})\, , \nonumber
\\
\Pv (\mathrm{7~TeV}) = 0.59 \pm 0.06 \, (\mathrm{stat.}) \, \pm 0.08 \, (\mathrm{syst.} ) \, \pm 0.010 \, (\mathrm{BR}) \, ^{+0.005}_{-0.003} (\mathrm{extr. }) \, , \nonumber
\end{eqnarray}
where uncertainties due to the extrapolation into the full phase space and branching ratios are negligible.
The values are compatible with the results from other experiments at different collision energies and for different colliding systems~\cite{CLEO,ALEPH,charmcdf,CDFPv,STARDmeson,aliceDmeson7TeV,ATLASDmeson}, as shown in Fig.~\ref{fig:pv}~(left)\footnote{
The ${\rm P}_{v}$ value of Ref.~\cite{CLEO} was corrected by the BR of Ref.~\cite{pdg} in Ref.~\cite{ADavid}. 
}. The weighted average of the experimental measurements reported in Ref.~\cite{ADavid}, with average $0.594 \pm 0.010$, and of the LHC data~\cite{aliceDmeson7TeV,ATLASDmeson} shown in Fig.~\ref{fig:pv}~is $\Pv^{\rm average} = 0.60 \pm 0.01$ which is represented by a solid yellow vertical band in the figure.

The expectation from na\"{\i}ve spin counting amounts to $\Pv^{\rm Spin \, counting} = 3/(3+1)=0.75$, showing a deviation from the data.
The argument of na\"{\i}ve spin counting originates from heavy-quark effective theory assuming large enough heavy-quark masses, leading to a negligible effect due to the mass difference between $\Dstar$ and $\Dplus$. 
In the PYTHIA 6.4.21~\cite{pythia} event generator the value for $\Pv$ is set by an input parameter (PARJ(13)) with a default value
 of $\Pv^{\rm Pythia} = 0.75$. We note, that there is only one parameter defining the probability that a charm or heavier meson has spin 1. 
Calculations combining the Lund symmetric fragmentation function with exact Clebsch-Gordan coefficient coupling from the virtual quark--antiquark pair to the final hadron state functions predicts $\Pv^{\rm Lund \, frag} \approx 0.63$~\cite{LundFrag} in good agreement with data. We note that in this model, due to the Clebsch-Gordan coefficient coupling, spin counting is automatic while differences in the hadron mass are taken into account in the fragmentation function by an exponential term. 
On the other hand, in the Statistical Model~\cite{shm-charm,stathadronization}, the ratio of the total yields of the directly formed charmed mesons $\Dstar$ to $\Dplus$, which have identical valence quark content, is expected to be 
$3 \cdot \left( m_{\Dstar} / m_{\Dplus} \right)^2  \cdot \exp\left(  -  ({m_{\Dstar} - m_{\Dplus} }) / T \right) \approx 1.4$ for a temperature parameter of $T=164$~MeV, 
where the factor of three comes from spin counting. 
We calculate $\Pv^{\rm Stat. \, Model} \approx 0.58 \pm 0.13 $ for $T=164\pm10$~MeV. 
Other implementations of the statistical model~\cite{becattini,rapp} predict similar values of $\Pv$, ranging between 0.55 and 0.64.
These $\Pv$ results are thus well described assuming either statistical hadronization of charm~\cite{shm-charm,stathadronization} or calculations considering the Lund symmetric fragmentation function~\cite{LundFrag}.

The weighted average of the total charm production cross section was calculated from the sum of  the total production cross section for $\Dzero$ and $\Dplus$ divided by the sum of their fragmentation ratios and the total production cross section for $\Dstar$ divided by its fragmentation ratio using the inverse of the squared statistical uncertainties as weights. The results are:
\begin{eqnarray}
\sigma_{\rm c\bar{c}}^{\rm tot}(\mathrm{2.76~TeV}) & = & 4.8 \pm 0.8  \, (\mathrm{stat.})  \, ^{+1.0}_{-1.3} \, (\mathrm{syst.}) \,
\pm 0.06 \, (\mathrm{BR})\,  \, \pm 0.1 \, (\mathrm{FF.})\,  \, \pm 0.1 \,(\mathrm{lum.})  \, ^{+2.6}_{-0.4} \, (\mathrm{extr.})~\mathrm{mb} \, ,
\label{eq:sigma-276} \nonumber
\\
\sigma_{\rm c\bar{c}}^{\rm tot}(\mathrm{7~TeV}) & = & 8.5 \pm 0.5  \, (\mathrm{stat.})  \, ^{+1.0}_{-2.4} \, (\mathrm{syst.}) \,
\pm 0.1 \, (\mathrm{BR})\,  \, \pm 0.2 \, (\mathrm{FF.})\,  \, \pm 0.3 \,(\mathrm{lum.})  \, ^{+5.0}_{-0.4} \, (\mathrm{extr.})~\mathrm{mb} \, .
\label{eq:sigma-7} \nonumber
\end{eqnarray}

The dependence of the total nucleon--nucleon charm production cross section~\cite{Lourenco,STAR,ATLASDmeson,LHCbDmeson,PHENIX} on the collision energy is shown in Fig.~\ref{fig:enercanvfull}~(right). The uncertainty boxes around the ATLAS~\cite{ATLASDmeson} and ALICE~\cite{aliceDmeson7TeV} points denote the extrapolation uncertainties alone, whilst the uncertainty bars are the overall uncertainties. Note that in case of proton--nucleus (pA) or deuteron--nucleus (dA) collisions, the measured cross sections have been scaled down by the number of binary nucleon--nucleon collisions  calculated in a Glauber model of the proton--nucleus or the deuteron--nucleus collision geometry. 
At $\sqrts=7$~TeV, our result and preliminary measurements by the ATLAS~\cite{ATLASDmeson} and the LHCb Collaboration~\cite{LHCbDmeson} are in fair agreement. 
The curves show the calculations at next-to-leading-order within the MNR framework~\cite{MNR} together with its uncertainties using the same parameters (and parameter uncertainties) mentioned before for FONLL. 
The dependence on the collision energy is described by pQCD calculations. 
We observe that all data points populate the upper band of the theoretical prediction. 

\section{Summary}
\label{sec:conclusions}
The measurement of the production of D mesons at mid-rapidity, $|y| < 0.5$, in \pp~collisions at $\sqrts=2.76~\tev$ from the hadronic decay channels $\DtoKpi$, $\DtoKpipi$, and $\DstartoDpi$, and their charge conjugates, has been reported.
The transverse momentum distributions are in agreement with pQCD calculations, 
even though the central prediction of FONLL~\cite{fonll,FONLLalice} (GM-VFNS~\cite{gmvfns,VFNSalice}) seems to underestimate~(overestimate) charm production. 
The $\pt$-differential cross sections are also in agreement with a rescaled reference computed from the cross section measured at a higher collision energy $\sqrt{s}=7~\tev$ with high statistics. The rescaling to the lower collision energy~\cite{EnergyExtrap} was performed by applying the collision energy dependence as computed by FONLL calculations. 
These two measurements, taken together, validate the $\sqrts$--scaling procedure and provide a reference for studying the QCD matter effects on charm quark production in Pb-Pb collisions at $\sqrts= 2.76$~TeV~\cite{aliceDmesonRaapaper}. 
An extrapolation to the full phase space using the shape of the distributions from FONLL yields the total production cross section of \ccbar pairs at LHC energies. 
The dependence on the collision energy is described by the pQCD expectations. 
The fraction of $\cdbar$ D mesons produced in a vector state is compatible with values from lower energies and different colliding systems.  

\newenvironment{acknowledgement}{\relax}{\relax}
\begin{acknowledgement}
\section*{Acknowledgements}
The ALICE collaboration would like to thank all its engineers and technicians for their invaluable contributions to the construction of the experiment and the CERN accelerator teams for the outstanding performance of the LHC complex. 
The ALICE Collaboration would like to thank M. Cacciari and H. Spiesberger for providing the pQCD predictions that are compared to these data. 
The ALICE collaboration acknowledges the following funding agencies for their support in building and running the ALICE detector:
Calouste Gulbenkian Foundation from Lisbon and Swiss Fonds Kidagan, Armenia;
Conselho Nacional de Desenvolvimento Cient\'{\i}fico e Tecnol\'{o}gico (CNPq), Financiadora de Estudos e Projetos (FINEP),
Funda\c{c}\~{a}o de Amparo \`{a} Pesquisa do Estado de S\~{a}o Paulo (FAPESP);
National Natural Science Foundation of China (NSFC), the Chinese Ministry of Education (CMOE)
and the Ministry of Science and Technology of China (MSTC);
Ministry of Education and Youth of the Czech Republic;
Danish Natural Science Research Council, the Carlsberg Foundation and the Danish National Research Foundation;
The European Research Council under the European Community's Seventh Framework Programme;
Helsinki Institute of Physics and the Academy of Finland;
French CNRS-IN2P3, the `Region Pays de Loire', `Region Alsace', `Region Auvergne' and CEA, France;
German BMBF and the Helmholtz Association;
General Secretariat for Research and Technology, Ministry of
Development, Greece;
Hungarian OTKA and National Office for Research and Technology (NKTH);
Department of Atomic Energy and Department of Science and Technology of the Government of India;
Istituto Nazionale di Fisica Nucleare (INFN) of Italy;
MEXT Grant-in-Aid for Specially Promoted Research, Ja\-pan;
Joint Institute for Nuclear Research, Dubna;
National Research Foundation of Korea (NRF);
CONACYT, DGAPA, M\'{e}xico, ALFA-EC and the HELEN Program (High-Energy physics Latin-American--European Network);
Stichting voor Fundamenteel Onderzoek der Materie (FOM) and the Nederlandse Organisatie voor Wetenschappelijk Onderzoek (NWO), Netherlands;
Research Council of Norway (NFR);
Polish Ministry of Science and Higher Education;
National Authority for Scientific Research - NASR (Autoritatea Na\c{t}ional\u{a} pentru Cercetare \c{S}tiin\c{t}ific\u{a} - ANCS);
Federal Agency of Science of the Ministry of Education and Science of Russian Federation, International Science and
Technology Center, Russian Academy of Sciences, Russian Federal Agency of Atomic Energy, Russian Federal Agency for Science and Innovations and CERN-INTAS;
Ministry of Education of Slovakia;
Department of Science and Technology, South Africa;
CIEMAT, EELA, Ministerio de Educaci\'{o}n y Ciencia of Spain, Xunta de Galicia (Conseller\'{\i}a de Educaci\'{o}n),
CEA\-DEN, Cubaenerg\'{\i}a, Cuba, and IAEA (International Atomic Energy Agency);
Swedish Research Council (VR) and Knut $\&$ Alice Wallenberg
Foundation (KAW);
Ukraine Ministry of Education and Science;
United Kingdom Science and Technology Facilities Council (STFC);
The United States Department of Energy, the United States National
Science Foundation, the State of Texas, and the State of Ohio.
\end{acknowledgement}

\newpage
%
\appendix
\section{The ALICE Collaboration}
\label{app:collab} 

\begingroup
\small
\begin{flushleft}
B.~Abelev\Irefn{org1234}\And
J.~Adam\Irefn{org1274}\And
D.~Adamov\'{a}\Irefn{org1283}\And
A.M.~Adare\Irefn{org1260}\And
M.M.~Aggarwal\Irefn{org1157}\And
G.~Aglieri~Rinella\Irefn{org1192}\And
A.G.~Agocs\Irefn{org1143}\And
A.~Agostinelli\Irefn{org1132}\And
S.~Aguilar~Salazar\Irefn{org1247}\And
Z.~Ahammed\Irefn{org1225}
% ...
\renewcommand\labelenumi{\textsuperscript{\theenumi}~}
\section*{Affiliation notes}
\renewcommand\theenumi{\roman{enumi}}
\begin{Authlist}
\item \Adef{M.V.Lomonosov Moscow State University, D.V.Skobeltsyn Institute of Nuclear Physics, Moscow, Russia}Also at: M.V.Lomonosov Moscow State University, D.V.Skobeltsyn Institute of Nuclear Physics, Moscow, Russia
\item \Adef{Institute of Nuclear Sciences, Belgrade, Serbia}Also at: "Vin\v{c}a" Institute of Nuclear Sciences, Belgrade, Serbia
\end{Authlist}
\section*{Collaboration Institutes}
\renewcommand\theenumi{\arabic{enumi}~}
\begin{Authlist}
\item \Idef{org1279}Benem\'{e}rita Universidad Aut\'{o}noma de Puebla, Puebla, Mexico
\item \Idef{org1220}Bogolyubov Institute for Theoretical Physics, Kiev, Ukraine
\item \Idef{org1262}Budker Institute for Nuclear Physics, Novosibirsk, Russia
\item \Idef{org1292}California Polytechnic State University, San Luis Obispo, California, United States
\item \Idef{org14939}Centre de Calcul de l'IN2P3, Villeurbanne, France
\item \Idef{org1197}Centro de Aplicaciones Tecnol\'{o}gicas y Desarrollo Nuclear (CEADEN), Havana, Cuba
\item \Idef{org1242}Centro de Investigaciones Energ\'{e}ticas Medioambientales y Tecnol\'{o}gicas (CIEMAT), Madrid, Spain
\item \Idef{org1244}Centro de Investigaci\'{o}n y de Estudios Avanzados (CINVESTAV), Mexico City and M\'{e}rida, Mexico
\item \Idef{org1335}Centro Fermi -- Centro Studi e Ricerche e Museo Storico della Fisica ``Enrico Fermi'', Rome, Italy
\item \Idef{org17347}Chicago State University, Chicago, United States
\item \Idef{org1288}Commissariat \`{a} l'Energie Atomique, IRFU, Saclay, France
\item \Idef{org1294}Departamento de F\'{\i}sica de Part\'{\i}culas and IGFAE, Universidad de Santiago de Compostela, Santiago de Compostela, Spain
\item \Idef{org1106}Department of Physics Aligarh Muslim University, Aligarh, India
\item \Idef{org1121}Department of Physics and Technology, University of Bergen, Bergen, Norway
\item \Idef{org1162}Department of Physics, Ohio State University, Columbus, Ohio, United States
\item \Idef{org1300}Department of Physics, Sejong University, Seoul, South Korea
\item \Idef{org1268}Department of Physics, University of Oslo, Oslo, Norway
\item \Idef{org1145}Dipartimento di Fisica dell'Universit\`{a} and Sezione INFN, Cagliari, Italy
\item \Idef{org1270}Dipartimento di Fisica dell'Universit\`{a} and Sezione INFN, Padova, Italy
\item \Idef{org1315}Dipartimento di Fisica dell'Universit\`{a} and Sezione INFN, Trieste, Italy
\item \Idef{org1132}Dipartimento di Fisica dell'Universit\`{a} and Sezione INFN, Bologna, Italy
\item \Idef{org1285}Dipartimento di Fisica dell'Universit\`{a} `La Sapienza' and Sezione INFN, Rome, Italy
\item \Idef{org1154}Dipartimento di Fisica e Astronomia dell'Universit\`{a} and Sezione INFN, Catania, Italy
\item \Idef{org1290}Dipartimento di Fisica `E.R.~Caianiello' dell'Universit\`{a} and Gruppo Collegato INFN, Salerno, Italy
\item \Idef{org1312}Dipartimento di Fisica Sperimentale dell'Universit\`{a} and Sezione INFN, Turin, Italy
\item \Idef{org1103}Dipartimento di Scienze e Tecnologie Avanzate dell'Universit\`{a} del Piemonte Orientale and Gruppo Collegato INFN, Alessandria, Italy
\item \Idef{org1114}Dipartimento Interateneo di Fisica `M.~Merlin' and Sezione INFN, Bari, Italy
\item \Idef{org1237}Division of Experimental High Energy Physics, University of Lund, Lund, Sweden
\item \Idef{org1192}European Organization for Nuclear Research (CERN), Geneva, Switzerland
\item \Idef{org1227}Fachhochschule K\"{o}ln, K\"{o}ln, Germany
\item \Idef{org1122}Faculty of Engineering, Bergen University College, Bergen, Norway
\item \Idef{org1136}Faculty of Mathematics, Physics and Informatics, Comenius University, Bratislava, Slovakia
\item \Idef{org1274}Faculty of Nuclear Sciences and Physical Engineering, Czech Technical University in Prague, Prague, Czech Republic
\item \Idef{org1229}Faculty of Science, P.J.~\v{S}af\'{a}rik University, Ko\v{s}ice, Slovakia
\item \Idef{org1184}Frankfurt Institute for Advanced Studies, Johann Wolfgang Goethe-Universit\"{a}t Frankfurt, Frankfurt, Germany
\item \Idef{org1215}Gangneung-Wonju National University, Gangneung, South Korea
\item \Idef{org1212}Helsinki Institute of Physics (HIP) and University of Jyv\"{a}skyl\"{a}, Jyv\"{a}skyl\"{a}, Finland
\item \Idef{org1203}Hiroshima University, Hiroshima, Japan
\item \Idef{org1329}Hua-Zhong Normal University, Wuhan, China
\item \Idef{org1254}Indian Institute of Technology, Mumbai, India
\item \Idef{org36378}Indian Institute of Technology Indore (IIT), Indore, India
\item \Idef{org1266}Institut de Physique Nucl\'{e}aire d'Orsay (IPNO), Universit\'{e} Paris-Sud, CNRS-IN2P3, Orsay, France
\item \Idef{org1277}Institute for High Energy Physics, Protvino, Russia
\item \Idef{org1249}Institute for Nuclear Research, Academy of Sciences, Moscow, Russia
\item \Idef{org1320}Nikhef, National Institute for Subatomic Physics and Institute for Subatomic Physics of Utrecht University, Utrecht, Netherlands
\item \Idef{org1250}Institute for Theoretical and Experimental Physics, Moscow, Russia
\item \Idef{org1230}Institute of Experimental Physics, Slovak Academy of Sciences, Ko\v{s}ice, Slovakia
\item \Idef{org1127}Institute of Physics, Bhubaneswar, India
\item \Idef{org1275}Institute of Physics, Academy of Sciences of the Czech Republic, Prague, Czech Republic
\item \Idef{org1139}Institute of Space Sciences (ISS), Bucharest, Romania
\item \Idef{org27399}Institut f\"{u}r Informatik, Johann Wolfgang Goethe-Universit\"{a}t Frankfurt, Frankfurt, Germany
\item \Idef{org1185}Institut f\"{u}r Kernphysik, Johann Wolfgang Goethe-Universit\"{a}t Frankfurt, Frankfurt, Germany
\item \Idef{org1177}Institut f\"{u}r Kernphysik, Technische Universit\"{a}t Darmstadt, Darmstadt, Germany
\item \Idef{org1256}Institut f\"{u}r Kernphysik, Westf\"{a}lische Wilhelms-Universit\"{a}t M\"{u}nster, M\"{u}nster, Germany
\item \Idef{org1246}Instituto de Ciencias Nucleares, Universidad Nacional Aut\'{o}noma de M\'{e}xico, Mexico City, Mexico
\item \Idef{org1247}Instituto de F\'{\i}sica, Universidad Nacional Aut\'{o}noma de M\'{e}xico, Mexico City, Mexico
\item \Idef{org23333}Institut of Theoretical Physics, University of Wroclaw
\item \Idef{org1308}Institut Pluridisciplinaire Hubert Curien (IPHC), Universit\'{e} de Strasbourg, CNRS-IN2P3, Strasbourg, France
\item \Idef{org1182}Joint Institute for Nuclear Research (JINR), Dubna, Russia
\item \Idef{org1143}KFKI Research Institute for Particle and Nuclear Physics, Hungarian Academy of Sciences, Budapest, Hungary
\item \Idef{org1199}Kirchhoff-Institut f\"{u}r Physik, Ruprecht-Karls-Universit\"{a}t Heidelberg, Heidelberg, Germany
\item \Idef{org20954}Korea Institute of Science and Technology Information, Daejeon, South Korea
\item \Idef{org1160}Laboratoire de Physique Corpusculaire (LPC), Clermont Universit\'{e}, Universit\'{e} Blaise Pascal, CNRS--IN2P3, Clermont-Ferrand, France
\item \Idef{org1194}Laboratoire de Physique Subatomique et de Cosmologie (LPSC), Universit\'{e} Joseph Fourier, CNRS-IN2P3, Institut Polytechnique de Grenoble, Grenoble, France
\item \Idef{org1187}Laboratori Nazionali di Frascati, INFN, Frascati, Italy
\item \Idef{org1232}Laboratori Nazionali di Legnaro, INFN, Legnaro, Italy
\item \Idef{org1125}Lawrence Berkeley National Laboratory, Berkeley, California, United States
\item \Idef{org1234}Lawrence Livermore National Laboratory, Livermore, California, United States
\item \Idef{org1251}Moscow Engineering Physics Institute, Moscow, Russia
\item \Idef{org1140}National Institute for Physics and Nuclear Engineering, Bucharest, Romania
\item \Idef{org1165}Niels Bohr Institute, University of Copenhagen, Copenhagen, Denmark
\item \Idef{org1109}Nikhef, National Institute for Subatomic Physics, Amsterdam, Netherlands
\item \Idef{org1283}Nuclear Physics Institute, Academy of Sciences of the Czech Republic, \v{R}e\v{z} u Prahy, Czech Republic
\item \Idef{org1264}Oak Ridge National Laboratory, Oak Ridge, Tennessee, United States
\item \Idef{org1189}Petersburg Nuclear Physics Institute, Gatchina, Russia
\item \Idef{org1170}Physics Department, Creighton University, Omaha, Nebraska, United States
\item \Idef{org1157}Physics Department, Panjab University, Chandigarh, India
\item \Idef{org1112}Physics Department, University of Athens, Athens, Greece
\item \Idef{org1152}Physics Department, University of Cape Town, iThemba LABS, Cape Town, South Africa
\item \Idef{org1209}Physics Department, University of Jammu, Jammu, India
\item \Idef{org1207}Physics Department, University of Rajasthan, Jaipur, India
\item \Idef{org1200}Physikalisches Institut, Ruprecht-Karls-Universit\"{a}t Heidelberg, Heidelberg, Germany
\item \Idef{org1325}Purdue University, West Lafayette, Indiana, United States
\item \Idef{org1281}Pusan National University, Pusan, South Korea
\item \Idef{org1176}Research Division and ExtreMe Matter Institute EMMI, GSI Helmholtzzentrum f\"ur Schwerionenforschung, Darmstadt, Germany
\item \Idef{org1334}Rudjer Bo\v{s}kovi\'{c} Institute, Zagreb, Croatia
\item \Idef{org1298}Russian Federal Nuclear Center (VNIIEF), Sarov, Russia
\item \Idef{org1252}Russian Research Centre Kurchatov Institute, Moscow, Russia
\item \Idef{org1224}Saha Institute of Nuclear Physics, Kolkata, India
\item \Idef{org1130}School of Physics and Astronomy, University of Birmingham, Birmingham, United Kingdom
\item \Idef{org1338}Secci\'{o}n F\'{\i}sica, Departamento de Ciencias, Pontificia Universidad Cat\'{o}lica del Per\'{u}, Lima, Peru
\item \Idef{org1316}Sezione INFN, Trieste, Italy
\item \Idef{org1271}Sezione INFN, Padova, Italy
\item \Idef{org1313}Sezione INFN, Turin, Italy
\item \Idef{org1286}Sezione INFN, Rome, Italy
\item \Idef{org1146}Sezione INFN, Cagliari, Italy
\item \Idef{org1133}Sezione INFN, Bologna, Italy
\item \Idef{org1115}Sezione INFN, Bari, Italy
\item \Idef{org1155}Sezione INFN, Catania, Italy
\item \Idef{org1322}Soltan Institute for Nuclear Studies, Warsaw, Poland
\item \Idef{org36377}Nuclear Physics Group, STFC Daresbury Laboratory, Daresbury, United Kingdom
\item \Idef{org1258}SUBATECH, Ecole des Mines de Nantes, Universit\'{e} de Nantes, CNRS-IN2P3, Nantes, France
\item \Idef{org1304}Technical University of Split FESB, Split, Croatia
\item \Idef{org1168}The Henryk Niewodniczanski Institute of Nuclear Physics, Polish Academy of Sciences, Cracow, Poland
\item \Idef{org17361}The University of Texas at Austin, Physics Department, Austin, TX, United States
\item \Idef{org1173}Universidad Aut\'{o}noma de Sinaloa, Culiac\'{a}n, Mexico
\item \Idef{org1296}Universidade de S\~{a}o Paulo (USP), S\~{a}o Paulo, Brazil
\item \Idef{org1149}Universidade Estadual de Campinas (UNICAMP), Campinas, Brazil
\item \Idef{org1239}Universit\'{e} de Lyon, Universit\'{e} Lyon 1, CNRS/IN2P3, IPN-Lyon, Villeurbanne, France
\item \Idef{org1205}University of Houston, Houston, Texas, United States
\item \Idef{org20371}University of Technology and Austrian Academy of Sciences, Vienna, Austria
\item \Idef{org1222}University of Tennessee, Knoxville, Tennessee, United States
\item \Idef{org1310}University of Tokyo, Tokyo, Japan
\item \Idef{org1318}University of Tsukuba, Tsukuba, Japan
\item \Idef{org21360}Eberhard Karls Universit\"{a}t T\"{u}bingen, T\"{u}bingen, Germany
\item \Idef{org1225}Variable Energy Cyclotron Centre, Kolkata, India
\item \Idef{org1306}V.~Fock Institute for Physics, St. Petersburg State University, St. Petersburg, Russia
\item \Idef{org1323}Warsaw University of Technology, Warsaw, Poland
\item \Idef{org1179}Wayne State University, Detroit, Michigan, United States
\item \Idef{org1260}Yale University, New Haven, Connecticut, United States
\item \Idef{org1332}Yerevan Physics Institute, Yerevan, Armenia
\item \Idef{org15649}Yildiz Technical University, Istanbul, Turkey
\item \Idef{org1301}Yonsei University, Seoul, South Korea
\item \Idef{org1327}Zentrum f\"{u}r Technologietransfer und Telekommunikation (ZTT), Fachhochschule Worms, Worms, Germany
\end{Authlist}
\endgroup

\end{document}